\documentclass{PoS}
\usepackage{epsfig,cite,color,axodraw}

\title{
\vspace*{-1.9cm}
\begin{minipage}{\textwidth}
{\normalfont\small LTH 1084, Nikhef 2016-023
\hspace{\fill} May 2016}\\
\end{minipage}\\[60pt]
 First Forcer results on deep-inelastic scattering \\ and related quantities}

\ShortTitle{Forcer results on DIS and related quantities}

\author{B. Ruijl$\,$\thanks
        {also at~ Leiden Centre of Data Science, Leiden University,
        Niels Bohrweg 1, 2333 CA Leiden, The Netherlands.},
        T. Ueda, J.A.M. Vermaseren \phantom {g} \\
        \mbox{Nikhef Theory Group, Science Park 105, 1098 XG Amsterdam,
        The Netherlands}\\
        E-mails: \email{benrl@nikhef.nl, tueda@nikhef.nl, t68@nikhef.nl}}

\author{J. Davies, \speaker{A. Vogt} \\ 
        \mbox{Department of Mathematical Sciences, University of Liverpool,
        Liverpool L69 3BX, UK}\\
        E-mails: \email{Joshua.Davies@liv.ac.uk, Andreas.Vogt@liv.ac.uk} 
        \\ \\ \\ }

\abstract{
We present results on the fourth-order splitting functions and coefficient
functions obtained \mbox{using} {\sc Forcer}, a four-loop generalization of the
{\sc Mincer} program for the parametric reduction of self-energy integrals. 
We have computed the respective lowest three even-$N$ and odd-$N$ moments for 
the non-singlet splitting functions and the non-singlet coefficient functions 
in electromagnetic and $\nu\!+\!\bar{\nu}$ charged-current deep-inelastic 
scattering, and the $N=2$ and $N=4$ results for the corresponding 
flavour-singlet quantities.  
Enough moments have been obtained for an LLL-based determination of the 
analytic $N$-dependence of the $\nft$ and $\nfs$ parts, respectively, of the 
singlet and non-singlet splitting functions.
The large-$N$ limit of the latter provides the complete $\nfs$ contributions 
to the four-loop cusp anomalous dimension. Our results also provide additional 
evidence of a non-vanishing contribution of quartic group invariants to the 
cusp anomalous dimension.
}

\FullConference{Loops and Legs in Quantum Field Theory\\
		24-29 April 2016, Leipzig, Germany}

\newcommand{\beq}{\begin{equation}}
\newcommand{\eeq}{\end{equation}}
\newcommand{\bea}{\begin{eqnarray}}
\newcommand{\eea}{\end{eqnarray}}
\newcommand{\nn}{\nonumber}

\newcommand{\gsim}{\raisebox{-0.05cm}{$\:\stackrel{>}{{\scriptstyle
 \sim}}\: $} }
\newcommand{\lsim}{\raisebox{-0.05cm}{$\:\stackrel{<}{{\scriptstyle
 \sim}}\: $} }

\newcommand{\ar}{a_{\rm s}}
\newcommand{\als}{\alpha_{\rm s}^{}}
\newcommand{\ass}{\alpha_{\rm s}}

\newcommand{\MSb}{$\overline{\mbox{MS}}$}

\def\as(#1){{\alpha_{\rm s}^{\:\!#1}}}

\def\z#1{{\zeta_{#1}^{}}}
\def\zs2{{\zeta_{2}^{\,2}}}
\def\zt2{{\zeta_{2}^{\,3}}}
\def\zf2{{\zeta_{2}^{\,4}}}
\def\ca{{C_{\!A}}}
\def\cas{{C^{\, 2}_{\!A}}}
\def\cat{{C^{\, 3}_{\!A}}}
\def\caf{{C^{\, 4}_{\!A}}}
\def\cf{{C_F}}
\def\nc{{n^{}_{\! c}}}
\def\nf{{n^{}_{\! f}}}
\def\nfs{{n^{\,2}_{\! f}}}
\def\nft{{n^{\,3}_{\! f}}}
\def\cfs{{C^{\, 2}_{\! F}}}
\def\cft{{C^{\, 3}_{\! F}}}

\def\dfAAna{{ {d_{\!A}^{\,abcd}\,d_{\!A}^{\,abcd} \over n_a} }}
\def\dfFAna{{ {d_{\!F}^{\,abcd}\,d_{\!A}^{\,abcd} \over n_a} }}
\def\dfRRna{{ {d_{\!F}^{\,abcd}\,d_{\!F}^{\,abcd} \over n_a} }}

\definecolor{Red}{rgb}{0.90,0.00,0.00}
\definecolor{Blue}{rgb}{0.00,0.00,1.00}
\definecolor{Green}{rgb}{0.1,0.40,0.05}

\def\S(#1){{{S}_{#1}}}
\def\Ss(#1,#2){{{S}_{#1,#2}}}
\def\Sss(#1,#2,#3){{{S}_{#1,#2,#3}}}
\def\Ssss(#1,#2,#3,#4){{{S}_{#1,#2,#3,#4}}}

\def\frkt#1#2{\mbox{\large{$\frac{#1}{#2}\:\!$}}}

\begin{document}

\section{Introduction}

\noindent
Impressive progress has been made in the past years on turning the 
next-to-next-to-leading order (NNLO, N$^2$LO) of perturbative QCD into the new 
default approximation for many hard processes, see, e.g., Refs.~\cite{NNLO} 
for some very recent calculations. 
While this accuracy is fully adequate for most quantities, there are cases 
where the next order, N$^3$LO, is of interest due to 
{\bf (a)} very high requirements on the theoretical accuracy, such as in the 
determination of the strong coupling constant~$\als$ from deep-inelastic 
scattering (DIS), see, e.g., Ref.~\cite{ABMas15}, 
or {\bf (b)} a slow convergence of the perturbation series, such as for Higgs 
production in proton-proton collisions, see, e.g., Refs.~\cite{Higgs}.

\vspace{1mm}
N$^3$LO analyses of processes with initial-state hadrons require, in principle, 
parton distributions $f_i^{}(x,\mu^2)$ determined at the same accuracy, 
including the renormalization-group dependence
\beq
\label{evol}
 \frac{d}{d \ln \mu^2} \, f_i^{}(x,\mu^2) 
 \; =\; \sum_{k}
 \left[ P^{}_{ik}(\als(\mu^2))_{\rm N^3LO}^{} 
 \otimes f_k^{}(\mu^2) \right](x)
\eeq
on the factorization and renormalization scale $\mu \equiv \mu_F^{} =\mu_R^{}$ 
with the splitting functions
\beq
\label{Pexp}
 P^{}_{ik}(x,\als)_{\rm N^3LO}^{}  \; =\;
          \als\,   P^{\,(0)}_{ik}(x) 
    \,+\, \as(2)\, P^{\,(1)}_{ik}(x)
    \,+\, \as(3)\, P^{\,(2)}_{ik}(x) 
    \,+\, \as(4)\, P^{\,(3)}_{ik}(x) 
 \; .
\eeq
Here $\otimes$ represents the Mellin convolution in the momentum fractions $x$,
and the sum over $k$ includes all $\nf$ effectively massless quark flavours; 
i.e., Eq.~(\ref{evol}) is a system of $(2\:\!\nf+\!1) \times (2\:\!\nf+\!1)$ 
coupled integro-differential equations. 
The splitting functions at NNLO \cite{Pnnlo} suggest that the effect of the
$\as(4)$ corrections in Eq.~(\ref{evol}) is very small at $x \,\gsim 
10^{\:\!-2}$, but this expectation cannot be extended with sufficient 
certainty to the full range of $x$ probed by benchmark processes at the LHC. 

\vspace*{1mm}
Here we report on the first steps of a project that aims to obtain a 
phenomenologically relevant amount of information on all functions 
$P^{\,(3)}_{ik}(x)$ in Eq.~(\ref{Pexp}). The idea is to employ the 
{\sc Forcer} program, see Ref.~\cite{tuLL2016}, to extend the 
{\sc Mincer}-based \cite{Mincer} fixed Mellin-$N$ calculations of 
Refs.~\cite{moms3loop} to four-loop accuracy, and then to construct 
approximate $x$-space expressions, analogous to those for 
$P^{\,(2)}_{ik}(x)$ in Refs.~\cite{NNLOappr}, from these results and 
information about the small-$x$ and large-$x$ limits 
\cite{smallx1,smallx2,smallx3,Korch89,DMS05,largex1,largex2}.

\vspace*{1mm}
Using basic symmetries, the system (\ref{evol}) can be decomposed into
$2\:\!\nf-\!1$ scalar equations and a $2 \!\times\! 2$~flavour-singlet system. 
The former (non-singlet) part consists of the $2 (\nf-\!1)$ flavour asymmetries
of quark-antiquark sums and differences, $q_i^{} \pm \bar{q}_i^{}$, and the 
total valence distribution, 
\beq
\label{qns}
  q_{{\rm ns},ik}^{\,\pm} \; = \; q_i^{} \pm \bar{q}_i^{} 
                           \,-\, (q_k^{} \pm \bar{q}_k^{}) 
\:\: , \quad 
  q_{\rm ns}^{\,\rm v} \; = \; 
  {\textstyle \sum_{\,r\,=1}^{\,\nf}} \, (q_r^{} - \bar{q}_r^{}) 
\eeq
with
\beq
\label{Pns}
  P_{\rm ns}^{\,\pm}   \; = \; P_{{\rm q}{\rm q}}^{\,\rm v}
    \pm P_{{\rm q}\bar{{\rm q}}}^{\,\rm v}
\; , \quad
  P_{\rm ns}^{\,\rm v} \; = \; P_{\rm qq}^{\,\rm v}
    - P_{{\rm q}\bar{{\rm q}}}^{\,\rm v} \:+\, \nf (P_{\rm qq}^{\,\sf s}
    - P_{{\rm q}\bar{{\rm q}}}^{\,\sf s}) 
  \; \equiv \; P_{\rm ns}^{\, -} + P_{\rm ns}^{\,\sf s} \; .
\eeq
Typical lowest-order diagrams for the different contributions in 
Eq.~(\ref{Pns}) are shown below.
 
\vspace*{6mm}\hspace*{1cm}
\begin{picture}(310,50)(-10,0)


\SetWidth{1.2}
\SetColor{OliveGreen}
\Photon(0,50)(10,40){1.5}{2}
\Photon(60,50)(50,40){1.5}{2}
\SetColor{Red}
\ArrowLine(0,0)(10,10)
\ArrowLine(10,10)(10,40)
\SetColor{OliveGreen}
\ArrowLine(10,40)(50,40)
\ArrowLine(50,40)(50,10)
\ArrowLine(50,10)(60,0)
\Gluon(10,10)(50,10){2.5}{6}

\Vertex(10,10){1.5}
\Vertex(50,10){1.5}
\Vertex(10,40){1.5}
\Vertex(50,40){1.5}

\SetWidth{0.8}
\SetColor{Black}
\DashLine(30,0)(30,50){5}


\SetWidth{1.2}
\SetColor{OliveGreen}
\Photon(80,50)(90,40){1.5}{2}
\Photon(140,50)(130,40){1.5}{2}
\SetColor{Red}
\ArrowLine(80,0)(90,10)
\Line(90,27)(90,40)
\SetColor{OliveGreen}
\ArrowLine(90,10)(130,10)
\ArrowLine(130,10)(140,0)
\Gluon(90,10)(90,27){2}{2}
\Gluon(130,10)(130,27){-2}{2}
\Line(90,40)(130,40)
\Line(130,27)(90,27)
\Line(130,40)(130,27)

\Vertex(90,10){1.5}
\Vertex(130,10){1.5}
\Vertex(90,40){1.5}
\Vertex(130,40){1.5}
\Vertex(90,27){1.5}
\Vertex(130,27){1.5}

\SetWidth{0.8}
\SetColor{Black}
\DashLine(110,0)(110,50){5}


\SetWidth{1.2}
\SetColor{OliveGreen}
\Photon(160,50)(170,40){1.5}{2}
\Photon(220,50)(210,40){1.5}{2}
\SetColor{Red}
\ArrowLine(160,0)(170,10)
\ArrowLine(170,40)(170,25)
\SetColor{OliveGreen}
\ArrowLine(170,10)(190,17.5)
\Line(190,17.5)(210,25)
\ArrowLine(210,10)(220,0)
\Gluon(170,10)(170,25){2}{2}
\Gluon(210,10)(210,25){-2}{2}
\ArrowLine(210,40)(170,40)
\Line(170,25)(187,18.5)
\ArrowLine(193,16.5)(210,10)
\ArrowLine(210,25)(210,40)

\Vertex(170,10){1.5}
\Vertex(210,10){1.5}
\Vertex(170,40){1.5}
\Vertex(210,40){1.5}
\Vertex(170,25){1.5}
\Vertex(210,25){1.5}

\SetWidth{0.8}
\SetColor{Black}
\DashLine(190,0)(190,50){5}


\SetWidth{1.2}
\SetColor{OliveGreen}
\Photon(240,50)(250,40){1.5}{2}
\Photon(300,50)(290,40){1.5}{2}
\SetColor{Red}
\ArrowLine(240,0)(250,10)
\Line(250,27)(250,40)
\SetColor{OliveGreen}
\Line(250,10)(262,10)
\ArrowLine(262,10)(290,10)
\ArrowLine(290,10)(300,0)
\Gluon(250,10)(250,27){2}{2}
\Gluon(262,10)(278,27){2}{3}
\Gluon(290,10)(290,27){-2}{2}
\Line(250,40)(290,40)
\Line(290,27)(250,27)
\Line(290,40)(290,27)

\Vertex(250,10){1.5}
\Vertex(290,10){1.5}
\Vertex(250,40){1.5}
\Vertex(290,40){1.5}
\Vertex(250,27){1.5}
\Vertex(290,27){1.5}
\Vertex(262,10){1.5}
\Vertex(278,27){1.5}

\SetWidth{0.8}
\SetColor{Black}
\DashLine(270,0)(270,50){5}

\end{picture}

\vspace{3mm}
$\hspace*{1.5cm}
P_{\rm qq}^{\,\rm v} = {\cal O}(\als) 
  \quad\quad\;\; 
P_{\rm qq}^{\,\sf s}, \, P_{{\rm q}\bar{{\rm q}}}^{\,\sf s}\,:\, \ass^2
  \qquad\quad
P_{{\rm q}\bar{{\rm q}}}^{\,\rm v}\,:\, \ass^2
  \qquad\quad
P_{{\rm q}\bar{{\rm q}}}^{\,\sf s} \neq P_{\rm qq}^{\,\sf s}\,:\, \ass^3$ 
 
\noindent
The remaining flavour-singlet quantities and their scale dependence (evolution)
are given by 
\beq
\label{qPsg}
  q^{\,}_{\,\sf s} \; = \; 
  {\textstyle \sum_{\:\!r=1}^{\,\nf}} \, (q_r^{} + \bar{q}_r^{})
\; , \quad
  \frac{d}{d \ln\mu^2} \,
  \bigg( \begin{array}{c} \! q^{}_{\:\!\sf s} \! \\[-0.5mm] \! g\! 
           \end{array} \bigg) 
  \; = \; \bigg( \! \begin{array}{cc} 
     \, P_{\rm qq} \, & P_{\rm qg} \, \\[-0.5mm]
     \, P_{\rm gq} \, & P_{\rm gg} \, \end{array} \! \bigg) \otimes 
  \bigg( \begin{array}{c} \!q^{}_{\,\sf s}\! \\[-0.5mm] \! g\! 
           \end{array} \bigg)
\; ,
\eeq
where $g(x,\mu^2)$, abbreviated by $g$, denotes the gluon distribution.
$P_{\rm qq}$ differs from $P_{\rm ns}^{\,+}$ in Eq.~(\ref{Pns}) by an 
additional pure singlet (ps) contribution starting at order $\as(2)$,
\beq
\label{Pqq} 
   P_{\rm qq} \: =\: P_{\rm ns}^{\,+} + \nf (P_{\rm qq}^{\:\rm s}
  + P_{\rm {q\bar{q}}}^{\:\rm s})
  \:\equiv\:  P_{\rm ns}^{\,+} + P_{\rm ps}^{} \; .
\eeq

Decompositions analogous to Eqs.~(\ref{Pns}) and (\ref{Pqq}) apply to the 
coefficient functions for inclusive DIS. 
In fact, following Refs.~\cite{moms3loop}, our calculations of the N$^3$LO 
splitting function are carried out via the unfactorized fourth-order 
coefficient functions in dimensional regularization, transformed to forward
amplitudes $A$ using the optical theorem and projected onto the $N$-th Mellin 
moment,
\beq
\label{Mellin}
  A(N) \; = \; \mbox{ \Large $\int$}_{\!\!0}^1 \, dx \; x^{\,N-1} A(x) \; ,
\eeq
by a dispersion relation in $x$. Like the operator-product expansion, this
approach determines either the even or the odd moments of the splitting and 
coefficient functions. 
Specifically, the even moments are obtained of quantities involving 
$q + \bar{q}$, such as $F_{2}$ and $F_{L}$ in electromagnetic and $\nu\!+\!
\bar{\nu}$ charged-current DIS, and the odd moments of quantities with
$q - \bar{q}$, such as $F_{3}$ in $\nu\!+\!\bar{\nu}$ charged-current DIS.
See Ref.~\cite{MRogal07} for a detailed discussion of these issues including 
the $\nu\!-\!\bar{\nu}$~cases.

\vspace*{1mm}
Before we turn to our new results, it is worthwhile to briefly recall the 
large-$N$ structure of the quark-quark splitting functions in the \MSb\ scheme 
employed throughout this article,
\beq
\label{largeN}
  \gamma_{\,\rm ns}^{\,(n)\pm,\rm v}(N) 
  \; \equiv \; - \, P_{\:\!\rm ns}^{\,(n)\pm,\rm v}(N) 
  \; = \; A_n \ln N 
  - B_n + C_n \, N^{-1}\ln N - D_n + {\cal O}_\pm ( N^{\,-2} ) \; .
\eeq
Here $A_n$ is the $(n\!+\!1)$-loop cusp anomalous dimension \cite{Korch89}, 
and $C_{n\,>\,2}$ has been predicted in terms of $A_{k\,<\,n}$ in 
Ref.~\cite{DMS05}. The differences between the $qq$ splitting functions are 
of order $N^{\,-2}$ at large~$N$.

\begin{figure}[h]
\vspace{-1mm}
\centerline{\hspace*{-2mm}\epsfig{file=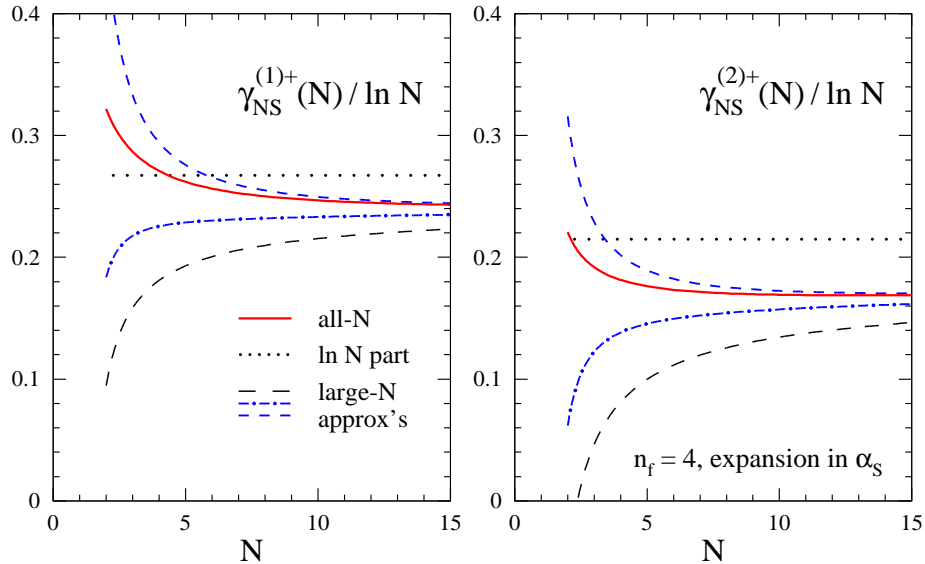,width=12.4cm,angle=0}}
\vspace{-2mm}
\caption{\label{Fig1}
The successive large-$N$ approximations in Eq.~(\protect\ref{largeN}) compared 
to the full NLO and NNLO results.$\!$
}
\vspace*{-4mm}
\end{figure}

\section{Low-$N$ results for splitting functions and coefficient functions}

\noindent
As an example of our analytic results, we present the $N=4$ anomalous dimension
$\gamma_{\,\rm gg}^{\,(3)}$, defined as in Eq.~(\ref{largeN}) above, for a 
general gauge group in terms of the expansion parameter 
$\ar \equiv \als/(4\:\! \pi)$,
{\small
\bea
\label{ggg4}
 \lefteqn{ \gamma_{\,\rm gg}^{\,(3)}(N\!=\!4) \:\:=\:\: 
    \textcolor{Blue}{ \caf } \* 
    \left( { 1502628149 \over 3375000 }
    + { 1146397 \over 11250 }\, \* \z3
    - { 504 \over 5 }\, \* \z5 \right)
  + \textcolor{Blue}{ \,\dfAAna } \*
    \left( { 21623 \over 150 }
\right. } \nn \\[-0.1mm] &&  \mbox{}\;\; \left.
    + { 15596 \over 15 }\, \* \z3
    - { 6048 \over 5 }\, \* \z5 \right)
  - \textcolor{Blue}{ \,\nf \, \* \cat \* } \left(
      { 20580892841 \over 72900000 }
    + { 12550223 \over 22500 }\, \* \z3
    - { 8613 \over 25 }\, \* \z4
    - { 4316 \over 27 }\, \* \z5 \right)
\nn \\[-0.1mm] &&  \mbox{}
  + \textcolor{Blue}{ \,\nf\, \* \dfFAna } \*
    \left( { 160091 \over 675 }
    + { 80072 \over 225 }\, \* \z3
    - { 48016 \over 45 }\, \* \z5 \right)
  - \textcolor{Blue}{ \nf \, \* \cas\, \* \cf \* } \left(
      { 4212122951 \over 41006250 }
\right. \nn \\[-0.1mm] &&  \mbox{}\;\; \left.
    - { 1170784 \over 5625 }\, \* \z3
    + { 418198 \over 1125 }\, \* \z4
    - { 17636 \over 45 }\, \* \z5 \right)
  + \textcolor{Blue}{ \,\nf \, \* \ca \* \cfs }\*
    \left( { 1913110089023 \over 26244000000 }
    + { 39313783 \over 101250 }\, \* \z3
\right. \nn \\[-0.1mm] &&  \mbox{}\;\; \left.
    + { 26741 \over 750 }\, \* \z4
    - { 3082 \over 5 }\, \* \z5 \right)
  + \textcolor{Blue}{ \,\nf \, \* \cft } \*
    \left( { 34764568601 \over 2099520000 }
    - { 958343 \over 40500 }\, \* \z3
    - { 18997 \over 2250 }\, \* \z4
    + { 908 \over 45 }\, \* \z5 \right)
\nn \\[-0.1mm] &&  \mbox{}
  - \textcolor{Blue}{ \,\nfs \, \* \cas \* } \left(
      { 3250393649 \over 218700000 }
    - { 2969291 \over 20250 }\, \* \z3
    + { 1566 \over 25 }\, \* \z4
    + { 1276 \over 135 }\, \* \z5 \right)
  - \textcolor{Blue}{ \,\nfs \, \* \cfs \* } \left(
      { 275622924731 \over 26244000000 }
\right. \nn \\[-0.1mm] &&  \mbox{}\;\; \left.
    - { 253369 \over 10125 }\, \* \z3
    + { 1078 \over 225 }\, \* \z4 \right)
  + \textcolor{Blue}{ \,\nfs \, \* \ca \* \cf } \*
    \left( { 136020246173 \over 3280500000 }
    - { 1672751 \over 10125 }\, \* \z3
    + { 15172 \over 225 }\, \* \z4 \right)
\nn \\[-0.1mm] &&  \mbox{}
  + \textcolor{Blue}{ \,\nfs\, \* \dfRRna }\*
    \left( { 75788 \over 675 }
    + { 3008 \over 15 }\, \* \z3
    - { 20416 \over 45 }\, \* \z5 \right)
  + \textcolor{Blue}{ \,\nft \, \* \cf \* }
    \left( { 1780699 \over 24300000 }
    - { 484 \over 675 }\, \* \z3 \right)
\nn \\[-0.1mm] &&  \mbox{}
  - \textcolor{Blue}{ \,\nft \, \* \ca \* } \left(
      { 20440457 \over 21870000 }
    - { 1888 \over 405 }\, \* \z3 \right)
\; .
\eea
}
Except for the last line \cite{LargeNf3}, Eq.~(\ref{ggg4}) is a new result.  
The complete set of fourth-order anomalous dimensions at $N \leq 4$ for 
Eq.~(\ref{qPsg}) and at $N \leq 6$ for Eq.~(\ref{Pns}) will be presented 
elsewhere \cite{RUVVprp}.

\vspace*{1mm}
Our results for $\gamma_{\,\rm ns}^{\,(3)\pm}(N)$ agree with the calculations 
at $N \leq 4$ in Refs.~\cite{P4ns1,VelizN2,VelizN34}. 
The numerical size of these quantities is shown in Fig.~2 for $\nf\!=\!3$ and 
$\nf\!=\!4$ light flavours. Taking into account the very slow large-$N$ 
convergence of $\gamma_{\,\rm ns}^{\,(n)}(N)/\ln N$ to $A_n$ in 
Eq.~(\ref{largeN}), see~Fig.~1 above, our results are consistent with, but not 
yet sufficient to improve on, the Pad\'e estimate of $A_3$ in Ref.~\cite{MVV7}.
Similarly, the $N$-dependent Pad\'e estimate used in N$^3$LO determinations of 
$\als$ from non-singlet DIS \cite{ABMas15} agrees with the calculated moments 
well within the large uncertainty assigned to it so far.

\begin{figure}[hbt]
\vspace{-2mm}
\centerline{\hspace*{-2mm}\epsfig{file=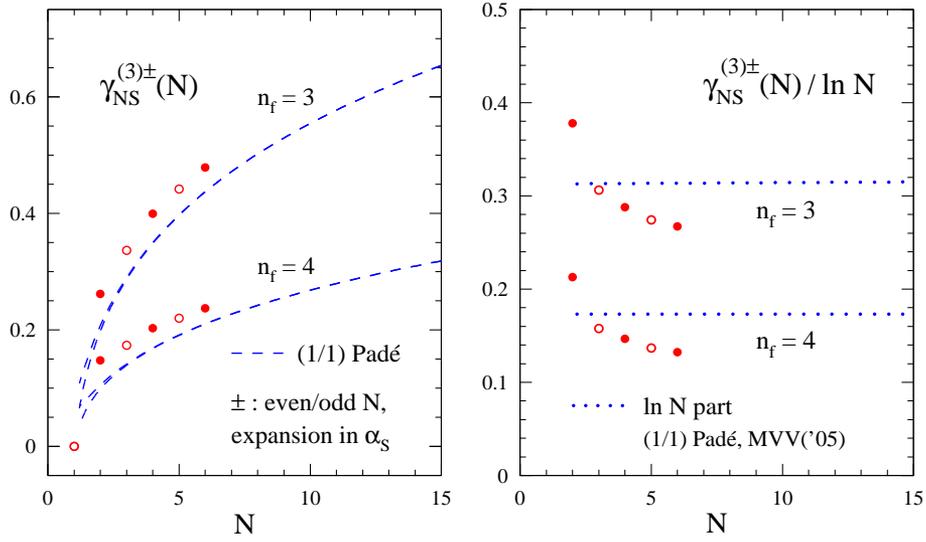,width=12.4cm,angle=0}}
\vspace{-2mm}
\caption{\label{Fig2}
The lowest three even-$N$ and odd-$N$ values, respectively, of the anomalous 
dimensions $\gamma_{\,\rm ns}^{\,(3)+}$ and $\gamma_{\,\rm ns}^{\,(3)-}$ in
Eqs.~(\protect\ref{Pns}) and (\protect\ref{largeN}), compared to Pad\'e 
estimates derived from the NNLO results of Ref.~\cite{Pnnlo}.
}
\vspace*{-1mm}
\end{figure}

\vspace*{3mm}
Inserting the QCD colour factors (the quartic group invariants are normalized 
as in Ref.~\cite{beta3}), the numerical expansions of the even-$N$ non-singlet 
anomalous dimensions at $\nf = 4$ are given by
\bea
\label{gns2num}
  \gamma_{\rm ns}^{\,+}(2,4) &=& 0.28294\, \als \left(
  1 + 0.7987\, \als 
    + 0.5451\, \as(2) 
    + 0.5215\, \as(3) 
    + \ldots\, \right)
\; , \nn \\[0.4mm]
  \gamma_{\rm ns}^{\,+}(4,4) &=& 0.55527\, \als \left(
  1 + 0.6851\, \als 
    + 0.4564\, \as(2) 
    + 0.3659 \, \as(3) 
    + \ldots\, \right)
\; , \\[0.4mm]
  \gamma_{\rm ns}^{\,+}(6,4) &=& 0.71645\, \als \left(
  1 + 0.6497\, \als 
    + 0.4368\, \as(2) 
   \textcolor{Blue}{\:+\, 0.3307 \, \as(3) } 
  + \ldots\, \right)
\nn \; .
\eea
%
%
%
The corresponding results for the odd-$N$ cases $\gamma_{\rm ns}^{\:\rm a}$ for 
$\,{\rm a} = -, {\rm v}\,$ are $\:\gamma_{\rm ns}^{\:\rm a}(1,\nf) \,=\, 0\:$, 
as required by fermion-number conservation, and
\bea
\label{gnsvnum}
  \gamma_{\,\rm ns}^{\:\rm a}(3,4) &=& 0.44210\, \als \left(
  1 + 0.7218\, \als
    + 0.4767\, \as(2) 
    + 0.3921\, \as(3) 
    + \ldots \right.
\nn \\[-0.2mm] & & \mbox{\hspace{3.3cm}} \left.
  +\, \delta_{\rm av} \left[ 0.0144\,\as(2) 
  \textcolor{Blue}{\:+\, 0.0045\,\as(3) } +\ldots \,\right]
  \right)
\; , \\[0.4mm]
  \gamma_{\,\rm ns}^{\:\rm a}(5,4) &=& 0.64369\, \als \left(
  1 + 0.6636\, \als
    + 0.4434\, \as(2) 
  \textcolor{Blue}{\:+\, 0.3421\,\as(3) } + \ldots \right.
\nn \\[-0.2mm] & & \mbox{\hspace{3.3cm}} \left.
  +\, \delta_{\rm av} \left[ 0.0032\,\as(2) 
  \textcolor{Blue}{\:+\, 0.0024\,\as(3) } + \ldots \, \right]
  \right)
\nn \; .
\eea
%

\noindent
The first two moments of the upper row of the splitting-function matrix in 
Eq.~(\ref{qPsg}) read
\bea
  \gamma_{\,\rm qq}^{}(2,4) &\,=\,& \phantom{-} 0.28294\, \als \left(
  1 + 0.6219\, \als 
    + 0.1461\, \as(2) 
\textcolor{Blue}{\:+\, 0.3662 \, \as(3) } 
    + \ldots \, \right)
\; , \nn \\
  \gamma_{\,\rm qq}^{}(4,4) &\,=\,& \phantom{-} 0.55527\, \als \left(
  1 + 0.6803\, \als 
    + 0.4278\, \as(2) 
\textcolor{Blue}{\:+\, 0.3459 \, \as(3) } 
    + \ldots \, \right)
\; , \\[1mm]
%
%
  \gamma_{\,\rm qg}^{}(2,4) &\,=\,& -0.21221\, \als \left(
  1 + 0.9004\, \als 
    - 0.1028\, \as(2) 
   \textcolor{Blue}{\:-\, 0.2367 \, \as(3) } 
    + \ldots\, \right)
\;, \nn \\
  \gamma_{\,\rm qg}^{}(4,4) &\,=\,& -0.11671\, \als \left(
  1 - 0.2801\, \als 
    - 0.9986\, \as(2) 
   \textcolor{Blue}{\:+\, 0.1297 \, \as(3) } 
    + \ldots\, \right)
\; . \eea
%
%
For the lower row we find
$\,\gamma_{\,\rm gi}(2,\nf) \,=\, - \gamma_{\,\rm qi}(2,\nf)\,$, 
as required by the momentum sum rule, and 
\bea
  \gamma_{\,\rm gq}^{}(4,4) &\,=\,& -0.07781\, \als \left(
  1 + 1.1152\, \als 
    + 0.8234\, \as(2) 
   \textcolor{Blue}{\:+\, 0.8833 \, \as(3) } 
    + \ldots\, \right)
\; , \nn \\
%
%
\label{ggg4num}
  \gamma_{\,\rm gg}^{}(4,4) &\,=\,& \phantom{-}1.21489\, \als \left(
  1 + 0.3835\, \als 
    + 0.1220\, \as(2) 
   \textcolor{Blue}{\:+\, 0.2406 \, \as(3) } 
    + \ldots\, \right)
\; . \eea
%
%
The relative N$^3$LO corrections are somewhat larger for $\nf= 3$, but are 
small in all cases with coefficients $\lsim \,1\,$ for $\nf\!=3,\ldots,6$ in 
Eqs.~(\ref{gns2num}) -- (\ref{ggg4num}), where our new results are given in 
blue.

\vspace*{1mm}
A check of the matrix in Eq.~(\ref{qPsg}) at $N>2$ is provided by a relation 
between the anomalous dimensions which emerges for $\nf = 1$ Majorana 
quarks and the choice   
$ \,\cf \,=\, 2\:\!T_F \,=\, \ca \,\equiv\, n_c\, \equiv\, n_{\rm colours} $ 
of the colour factors that leads to a supersymmetric theory \cite{AntFlo81}: 
The combination\footnote
{Up to NNLO, the same results are obtained by keeping the QCD value 
$T_F =1/2$ and setting $\nf = n_c\:\!$.}
\beq
  \Delta_{\,\rm S}^{(n)}(N) \,=\, \mbox{}
  - \gamma_{\,\rm qq}^{\,(n)}(N) - \gamma_{\,\rm gq}^{\,(n)}(N) 
  + \gamma_{\,\rm qg}^{\,(n)}(N) + \gamma_{\,\rm gg}^{\,(n)}(N)
\eeq
is supposed to vanish for a regularization that does not violate the 
supersymmetry. In dimensional regularization $\Delta_{\,\rm S}^{(n)}$ 
does not vanish, but is much simpler than the anomalous dimensions, see 
Ref.~\cite{AMV11} for a brief discussion at NNLO. We find that this 
expected simplification occurs also at N$^3$LO (at $N=4$, for now:
$\Delta_{\,\rm S}$ vanishes at $N=2$ already in QCD due to the momentum 
sum rule)~at
\beq
      (2 \nf)^2 \, \frac{d^{\,(4)}_{F\!F}}{n_a}
 \,=\, 2 \nf \, \frac{d^{\,(4)}_{FA}}{n_a}
 \,=\, 2 \nf \, \frac{d^{\,(4)}_{F\!F}}{n_c} 
 \,=\, \frac{d^{\,(4)}_{FA}}{n_c} 
 \,=\, \frac{d^{\,(4)}_{AA}}{n_a} 
 \;\; , \qquad
 d^{\,(4)}_{\rm xy} \equiv d_x^{\,abcd} d_y^{\,abcd}
\eeq
for the quartic group invariants with all particles in the adjoint 
representation. The additional factor of two for each power of $\nf$ 
in the QCD results is due to the transition to Majorana fermions.

\vspace{1mm}
We now briefly turn to the coefficient functions $C_a$ for DIS in massless 
perturbative QCD \cite{mvvC23L}; see Refs.~\cite{DISmass} for the important 
heavy-quark contributions. 
The size of the fourth-order corrections is illustrated in Fig.~3 for 
the structure functions $F_{2,\rm ns}$, $F_3$ and $F_{L,\rm ns}$ in 
charged-current $\nu\!+\!\bar{\nu}$~DIS. 
For~$F_2$ and $F_3$ the $\,\ln^8\! N\ldots \ln^2\! N\,$ large-$N$ contributions
to $C_{a,q}^{\,(4)}(N)$ are fixed by the soft-gluon exponentiation~\cite{MVV7},
and~the subleading $N^{-1} (\ln^7\! N \ldots \ln^4\! N)$ terms by the 
double-logarithmic resummations in Refs.~\cite{largex3,largex1}. 
For $C_{L,q}^{\,(4)}(N)$ the latter provide the 
$N^{-1} (\ln^6\! N \ldots \ln^4\! N)$ contributions.

\begin{figure}[hbt]
\centerline{\hspace*{-2mm}\epsfig{file=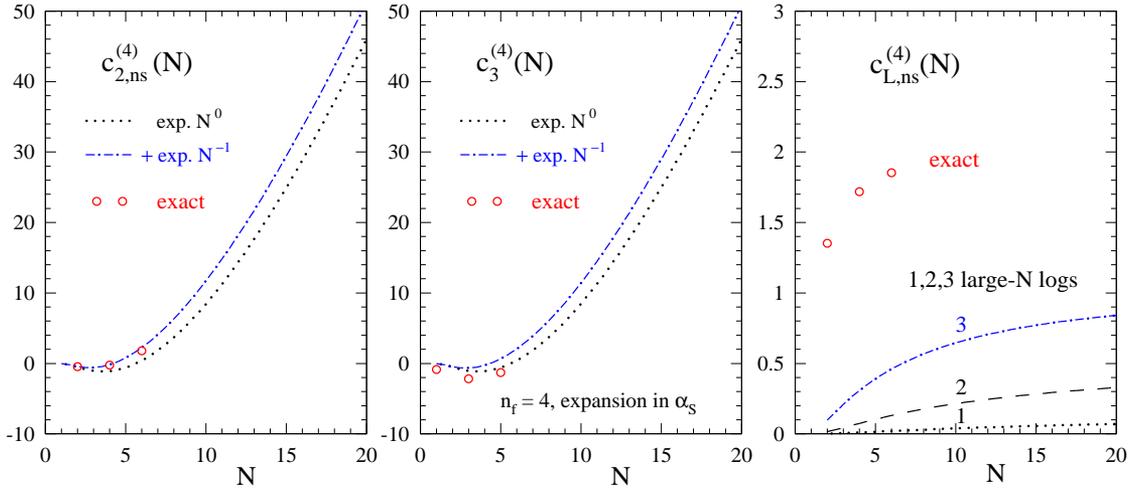,width=15cm,angle=0}}
\vspace{-2mm}
\caption{\label{Fig3}
The moments calculated so far of the fourth-order coefficient functions 
$c_{2,\rm ns\,}^{\,(4)}$, $c_3^{\,(4)}$ and $c_{L,\rm ns}^{\,(4)}$ for
$\nu\!+\!\bar{\nu}$ charged-current DIS at $\nf=4$. 
Also show are the contributions provided by large-$N$ resummations.
}
\end{figure}

\noindent
The numerical $\als$ expansions of these coefficient functions at low values 
of $N$ read, for $\nf=4$, 
\bea
  C_{2,\rm ns}(2,4) &\,=\,&
  1 + 0.0354\, \als - 0.0231\, \as(2) - 0.0613\, \as(3)
\textcolor{Blue}{ \,-\, 0.4746\, \as(4) }
    + \ldots \;
\; , \nn \\[0.2mm]
\label{c2exp}
  C_{2,\rm ns}(4,4) &\,=\,& 
  1 + 0.4828\, \als + 0.4711\, \as(2) + 0.4727\, \as(3) 
\textcolor{Blue}{ \,-\, 0.2458\, \as(4) }
    + \ldots \;
\; , \\[0.2mm]
  C_{2,\rm ns}(6,4) &\,=\,& 
  1 + 0.8894\, \als + 1.2053\, \as(2) + 1.7571\, \as(3) 
\textcolor{Blue}{ \,+\, 1.7748\, \as(4) }
    + \ldots \;
\; , \nn
\\[2mm] 
%
%
%
%
  C_{3,\rm ns}(1,4) &\,=\,&
  1 - 0.3183\, \als - 0.3293\, \as(2) - 0.4467\, \as(3)
    - 1.0512\, \as(4)
    + \ldots
\nn \\[-0.2mm] & & \mbox{\hspace{3.3cm}}
  +\, \delta_{\rm av} \left[ 0.0533\,\as(3)
  + 0.1999\,\as(4) +\ldots \,\right]
\; , \nn \\[0.2mm]
\label{c3exp}
  C_{3,\rm ns}(3,4) &\,=\,&
  1 + 0.1326\, \als - 0.0852\, \as(2) - 0.5202\, \as(3)
\textcolor{Blue}{ \,-\, 2.2510\, \as(4) }
    + \ldots
\nn \\[-0.2mm] & & \mbox{\hspace{3.3cm}}
  +\, \delta_{\rm av} \left[ 0.0202\,\as(3)
  \textcolor{Blue}{\:+\, 0.0805\,\as(4) } +\ldots \,\right]
\; , \\[0.2mm]
  C_{3,\rm ns}(5,4) &\,=\,&
  1 + 0.6166\, \als + 0.6042\, \as(2) + 0.4214\, \as(3)
\textcolor{Blue}{ \,-\, 1.3217\, \as(4) }
    + \ldots
\nn \\[-0.2mm] & & \mbox{\hspace{3.13cm}}
  +\, \delta_{\rm av} \left[ 0.00788\,\as(3)
  \textcolor{Blue}{\:+\, 0.0422\,\as(4) } +\ldots \,\right]
\; , \nn
\\[2mm]
%
%
%
%
  C_{L,\rm ns}(2,4) &\,=\,& 0.14147\, \als \left(
  1 + 1.7270\, \als + 3.7336\, \as(2) 
  \textcolor{Blue}{\:+\, 9.5619 \, \as(3) } 
  + \ldots \,\right)
\; , \nn \\[0.5mm]
\label{cLexp}
  C_{L,\rm ns}(4,4) &\,=\,& 0.08488\, \als \left(
  1 + 2.5619\, \als + 6.9208\, \as(2) 
  \textcolor{Blue}{\:+\, 20.251 \, \as(3) } 
  + \ldots \,\right)
\; , \\[0.5mm]
  C_{L,\rm ns}(6,4) &\,=\,& 0.06063\, \als \left(
  1 + 3.1557\, \als + 9.6370\, \as(2) 
  \textcolor{Blue}{\:+\, 30.572 \, \as(3) } 
  + \ldots \,\right)
\; . \nn 
\eea
%
%
%
%
The first moment of $F_3$ is the Gross--Llewellyn-Smith (GLS) sum rule; its
coefficients in Eq.~(\ref{c3exp}) agree with those of Refs.~\cite{SumRule}, 
where the $\delta_{\rm av}$ part [cf.~Eq.~(\ref{gnsvnum})] is called the 
singlet contribution.

\section{Large-$\nf$ all-$N$ parts of N$^3$LO splitting functions \& 
the cusp anomalous dimension}
 
\noindent
The extension of our above results to higher values of $N$ will require very
considerable computing resources and further optimizations of our programs.
The situation is more favourable for some leading and subleading large-$\nf$
contributions, which do not involve the hardest diagram topologies. 
For example, the top-level diagrams contributing to the $\nfs$ parts of the 
anomalous dimensions $\gamma^{\,(3)\pm}_{\,\rm ns}$ are the same as for the
$\nf$ parts of the NNLO contributions $\gamma^{\,(2)\pm}_{\,\rm ns}$ in 
Ref.~\cite{MVV2}, 
 
\vspace*{-2.5mm}
\hspace*{4mm}
\includegraphics[bb = 130 110 310 710, scale = 0.3, angle = 270, clip]{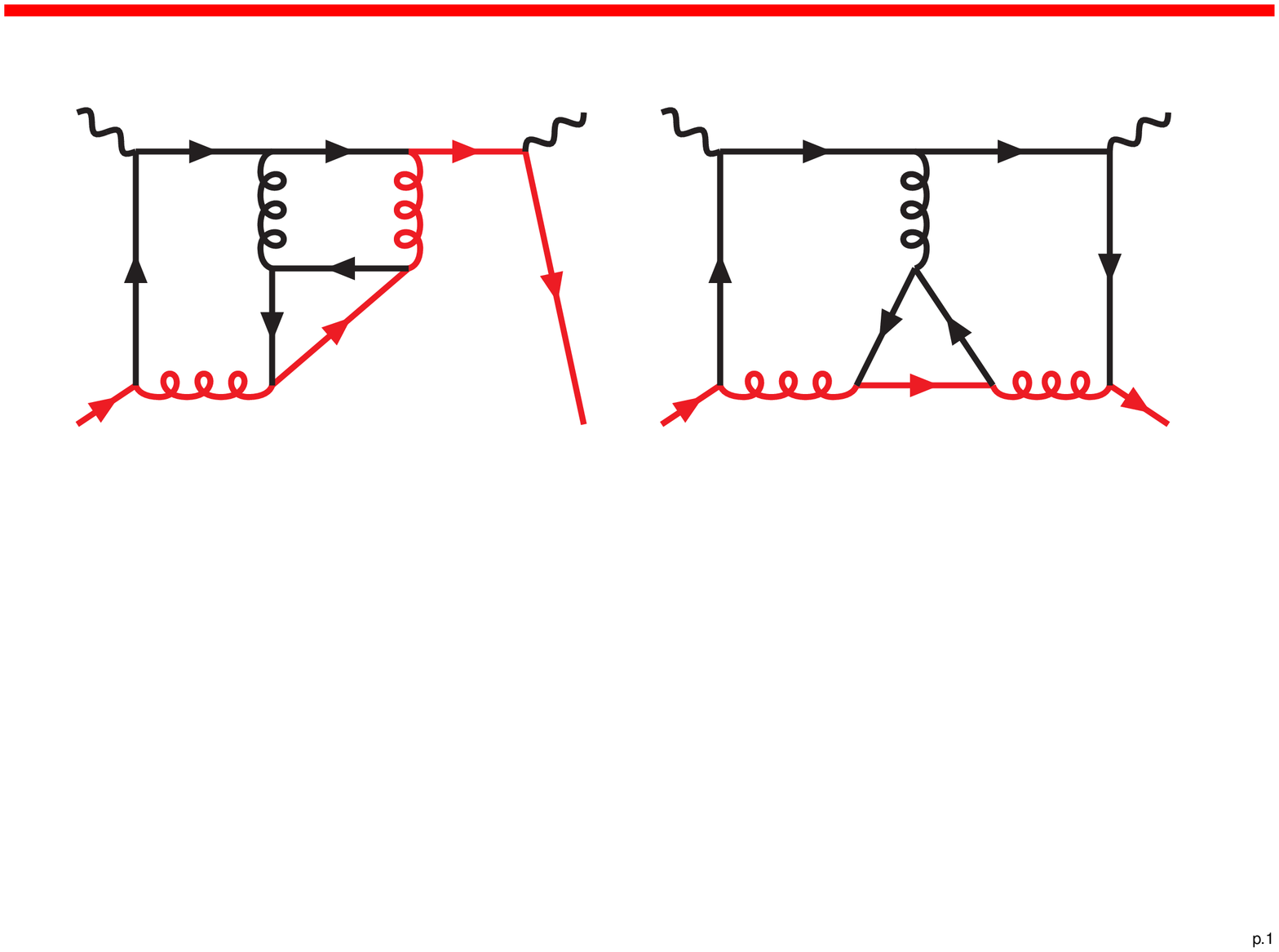}
\includegraphics[bb = 130 110 310 710, scale = 0.3, angle = 270, clip]{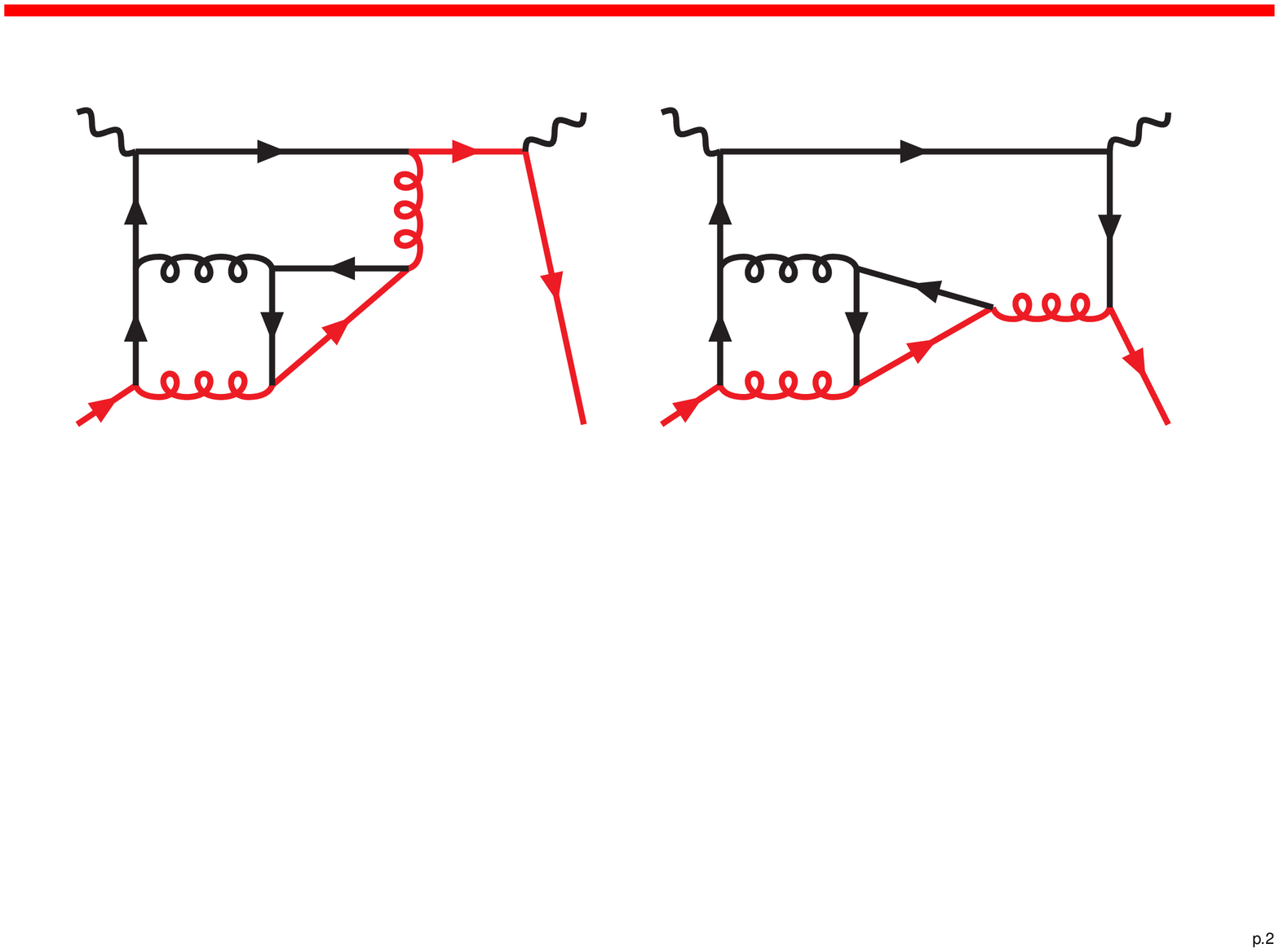}

\vspace*{2.5mm}
\noindent
but with an additional quark loop inserted into one of the gluon propagators.
These diagrams have the colour factor $\cf \ca \nfs$; the $\cfs\, \nfs$ cases
are even simpler, as is the $\cf \nft$ part derived in Ref.~\cite{LargeNf1}.

\vspace*{1mm}
It is convenient to write the colour-factor decomposition the $\nfs$
parts of $\gamma^{\,(3)\pm}_{\,\rm ns}$ in two ways,
\bea
\label{gnsNf2}
  \gamma^{\,(3)\pm}_{\,\rm ns}(N) \Big|_{\,\nfs} &\,=\,& 
  \cf \nfs\, \left\{ \,\cf \; 2A(N) \hspace*{0.65cm} 
  + (\ca-2\cf) B_\pm(N) \right\}
\nn \\[-2mm] &=& 
  \cf \nfs\, \left\{ \,\cf (2A(N)-2B_\pm(N) ) + \; \ca\, B_\pm(N) \right\}
\; .
\eea
$A(N)$ is the large-$n_c$ part; it is the same for the even-$N$ ($+$) and 
odd-$N$ ($-$) cases and should include only non-alternating harmonic sums 
\cite{HSums}. Once $A(N)$ is known, it is possible to determine $B_+(N)$ and
$B_-(N)$ from the $\cf$ parts in the second line of Eq.~(\ref{gnsNf2}) which
require only two-loop diagrams.
 
\vspace*{1mm}
We have computed the even and odd moments up to $N = 20$ for the determination 
of $A(N)$, and the even-$N$ or odd-$N$ moments up to $N = 42$ for $B_+(N)$ and
$B_-(N)$, respectively. These calculations are sufficient to determine all 
three function using an LLL-based program \cite{LLLprg}, see also 
Refs.~\cite{LLLappl}, with a sufficient number of validation constraints. 
The resulting large-$n_c$ contribution reads
\bea
\label{gns3nf2a}
  && \hspace*{-4mm} \gamma^{\,(3)}_{\,\rm ns}(N) |_{\cf\nc\nfs}
  \; = \;
          \frkt{127}{18}
        + \frkt{1}{81} \* \Big(
             \frkt{20681}{2}\, \* \eta
           + 2119\, \* \S(1)
           - 2275\, \* \eta^2
           - 20460\, \* D_{1}^{2}
           + 3392\, \* \S(1) \* \eta
           - 5036\, \* \S(2)
           \Big)
\nn \\[-0.6mm] & & \mbox{} 
        + \frkt{4}{81} \* \big(
             118\, \* \eta^3
           - 886\, \* D_{1}^{3}
           - 914\, \* \S(1) \* \eta^2
           - 848\, \* \S(1) \* D_{1}^{2}
           - 152\, \* \Ss(1,2)
           - 416\, \* \S(2) \* \eta
           - 152\, \* \Ss(2,1)
           + 1148\, \* \S(3)
           \big)
\nn \\[0.2mm] & & \mbox{} 
        + \frkt{8}{27} \* \big(
           - 57\, \* D_{1}^{4}
           + 18\, \* \S(1) \* \eta^3
           - 24\, \* \S(1) \* D_{1}^{3}
           + 2\, \* \S(2) \* \eta^2
           + 128\, \* \S(2) \* D_{1}^{2}
           - 8\, \* \S(3) \* \eta
           + 40\, \* \Ss(1,3)
           + 80\, \* \Ss(2,2)
\nn \\[0.2mm] & & \mbox{} 
           + 120\, \* \Ss(3,1)
           - 159\, \* \S(4)
           \big)
        + \frkt{8}{9} \* \big(
           - 6\, \* \eta^5
           - 12\, \* D_{1}^{5}
           + 10\, \* \S(1) \* \eta^4
           - 24\, \* \S(1) \* D_{1}^{4}
           + 8\, \* \S(2) \* \eta^3
           + 4\, \* \S(3) \* \eta^2
\nn \\[0.2mm] & & \mbox{} 
           - 8\, \* \S(3) \* D_{1}^{2}
           + 4\, \* \Ss(3,1) \* \eta
           - 8\, \* \Sss(1,3,1)
           + 4\, \* \Ss(1,4)
           - 8\, \* \Ss(2,3)
           - 16\, \* \Ss(3,2)
           - 2\, \* \S(4) \* \eta
           - 20\, \* \Ss(4,1)
           + 24\, \* \S(5)
           \big)
\nn \\[0.2mm] & & \mbox{} 
        + \z3 \* \big\{ 
           - \frkt{44}{3}
           - \frkt{160}{9}\,\* ( \eta - 2\,\* \S(1) )
           + \frkt{16}{3}\, \* ( \eta^2 - 2\,\* D_{1}^{2} - 2\,\* \S(2) )
           \big\}    
        + \z4 \* \big\{
             12
           + 8\,\* \eta
           - 16\,\* \S(1)
           \big\}    
\; ,
\eea
where all sums are taken at $N$ and we have used the abbreviations
$D_i = (N\!+\!i)^{-1\!}$ and $\eta = D_0\! -\! D_1$.

\vspace*{1mm}
The large-$N$ limit (\ref{largeN}) of Eq.~(\ref{gns3nf2a}), together with the 
corresponding expressions for $B_\pm(N)$ \cite{DRUVVprp} yields the complete
$\nfs$ contribution to the four-loop quark cusp anomalous dimension,
\bea
\label{cusp4nf2}
   \gamma^{\,(3)}_{\,\rm cusp} 
  &\:=\:& \ldots 
     \,+\, \cf \* \ca \* \nfs \* \left(
             \frac{923}{81}
           - \frac{608}{81}\,\* \z2
           + \frac{2240}{27}\,\* \z3
           - \frac{112}{3}\,\* \z4
           \right)
\nn \\[0.5mm] && \mbox{} 
     \,+\, \cfs \* \nfs \* \left(
             \frac{2392}{81} 
           - \frac{640}{9}\,\* \z3 
           + 32\,\* \z4
           \right) 
         - \cf \* \nft \* \left(
             \frac{32}{81}
           - \frac{64}{27}\,\* \z3
           \right)
\; .
\eea

\begin{figure}[thb]
\vspace*{-1mm}
\centerline{\hspace*{-2mm}\epsfig{file=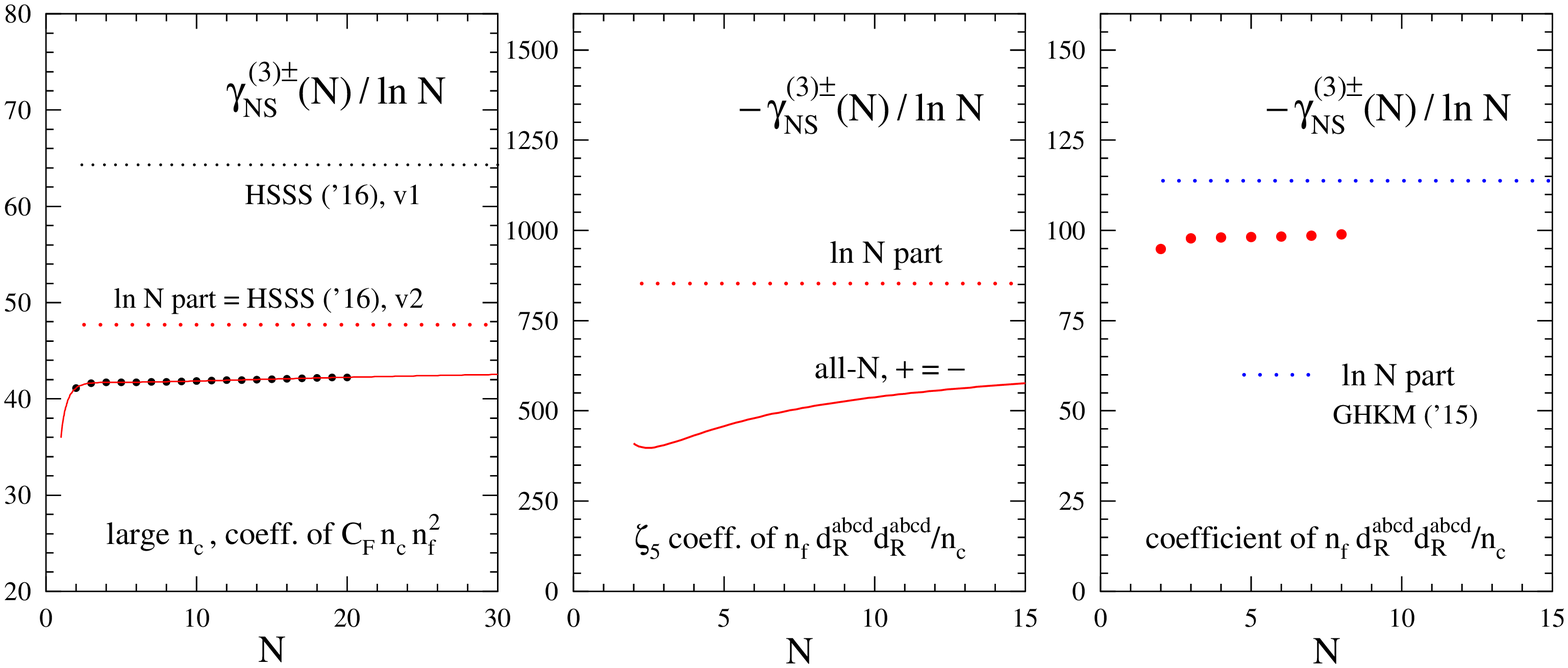,width=15cm,angle=0}}
\vspace{-2mm}
\caption{\label{Fig4}
Fermionic contributions to the N$^3$LO anomalous dimensions
$\gamma^{\,(3)\pm}(N)$, compared to their calculated (left two panels) and
conjectured (right panel) large-$N$ limits given by the respective parts of
$\gamma^{\,(3)}_{\,\rm cusp}$.
}
\vspace{-1mm}
\end{figure}

\noindent
The large-$n_c$ limit of Eq.~(\ref{cusp4nf2}) agrees with the second version
of Ref.~\cite{HSSS16}, in which an error has been fixed after we pointed out
a discrepancy with our result, see the left part of Fig.~4. 
The~$\cfs\, \nfs$ coefficient agrees with the result of Ref.~\cite{GHKM15}, 
which was converted to our notation and compared during this conference. 
The $\nft$ part of $\gamma^{\,(3)}_{\,\rm ns}(N)$, and hence the $\nft$ 
coefficient in Eq.~(\ref{cusp4nf2}), agrees with Ref.~\cite{LargeNf1}. 
Our results also agree with the prediction of the $N^{-1} \!\ln N$ coefficient 
in Ref.~\cite{DMS05} and the small-$x$ resummation result for the $1/N^5$ 
contribution \cite{smallx3}.

\vspace{2mm}
In the flavour singlet case, at least for the time being, only the $ \nft$ 
leading large-$\nf$ contributions can be determined in this manner; the results 
will be presented in Ref.~\cite{DRUVVprp}.

\vspace{2mm}
The question of whether or not the quartic group invariants contribute to the
four-loop cusp anomalous dimension has attracted some interest, see, e.g.,
Refs.~\cite{d4cusp}. The presence of such contributions would violate the
Casimir scaling, $\gamma_{\,\rm cusp, q} = \cf/\ca\, \gamma_{\,\rm cusp, g}$,
observed up to NNLO \cite{Pnnlo}. 
In our calculations, the relatively easiest contribution of this type is the  
$\,\nf\, d_{\!F}^{\,abcd} d_{\!F}^{\,abcd} / n_c $ part of the quark case, 
which appears as the corresponding $\ln N$ coefficient of 
$\gamma^{\,(3)}_{\,\rm ns}(N)$.
So far we have extended the calculation of this contribution to $N=8$.
Except for the $\z5$ part,
\beq
\label{cuspRRz5}
  \gamma^{\,(3)}_{\,\rm ns}(N)\bigg|_{\,\z5\,\nf d_{F\!F}^{(4)}/n_c} \;=\;
  \frac{1280}{3}\, \* \bigg[ \,
      2\,\*\S(1)(N) 
    - 3 
    + 17\,\* \bigg( \frac{1}{N} - \frac{1}{N+1} \bigg)
    - 6\, \*  \bigg( \frac{1}{N^2} + \frac{1}{(N+1)^2} \bigg)
  \bigg]
\; , 
\eeq
(the corresponding result in Ref.~\cite{VelizN34} is unfortunately incorrect
-- only four moments were available there, and the $1/N^2$ and $1/(N+1)^2$ 
contributions were erroneously assumed to be absent)
this is not sufficient for a determination of the all-$N$ result from which
$\gamma_{\,\rm cusp}^{\,(3)}$ can be read off. 
Together with the prime content of the denominators of the calculated moments,
Eq.~(\ref{cuspRRz5}) is suggestive, but not a positive proof, of a 
non-vanishing $\,\z5 \,\nf\, d_{\!F}^{\,abcd} d_{\!F}^{\,abcd} / n_c $
to $\gamma_{\,\rm cusp}^{\,(3)}$.  Moreover the calculated moments,
shown in Fig.~4, clearly point to a non-vanishing value; in particular, they 
are consistent with the numerical value proposed in Ref.~\cite{GHKM15} on the
basis of a conjectured relation to the quark-antiquark potential calculated in 
Refs.~\cite{Vqqbar} for this colour factor.

\section{Summary and outlook}

\noindent
We have presented the first computations of anomalous dimensions and 
coefficient functions at order $\as(4)$ with {\sc Forcer}, a new {\sc Form} 
\cite{FORM} program for the analytic evaluation of four-loop massless 
propagator integrals. Our results agree with those of all comparable 
calculations performed so far. Together with the calculation of the four-loop 
gluon propagator in the background gauge to all powers of the gauge parameter, 
this provides a robust validation of the {\sc Forcer} package.

\vspace{0.5mm}
So far we have extended previous calculations \cite{P4ns1,VelizN2,VelizN34}
of the non-singlet splitting functions for the evolution of the parton 
distributions of the proton by one moment each for $P_{\,\rm ns}^{\,(3)+}$
and $P_{\,\rm ns}^{\,(3)-}$. We have performed the first calculations, at 
$N=2$ and $N=4$, of the corresponding flavour-singlet quantities, and the
first calculations of fourth-order coefficient functions in DIS beyond
$N=1$~\cite{SumRule}.
The full results will be presented in Ref.~\cite{RUVVprp}, together with the
four-loop contributions to the renormalization factors $Z_5$ and $Z_A$
required if the Larin scheme for $\gamma_5$ \cite{gamma5} is used in the 
calculations. 

\vspace{0.5mm}
Unlike the four-loop renormalization of QCD, the calculations of moments of
structure functions require very considerable computing resources. 
Much more than thousand times the time of the third-order computation is 
required at $N=4$, and the scaling of the hardest topologies with $N$ is, at 
least so far, much worse than that of the {\sc Mincer} program in its final 
highly optimized form.

\vspace*{0.5mm}
Nevertheless, already now we have been able to calculate enough moments for
the determination of the all-$N$ expressions of the $\nfs$ contributions to
$\gamma_{\,\rm ns}^{\,(3)\pm}$ and the leading large-$\nf$ contributions to
their flavour singlet counterparts $\gamma_{\,\rm ik}^{\,(3)}$ via Diophantine
equations for the coefficients of the harmonic sums.
We do not expect that the determination of all-$N$ expressions in this manner 
can be extended far beyond the point we have reached now. 
However, we hope to be able to obtain more moments in the future, and to 
provide approximate results for the N$^3$LO splitting functions 
$P_{\rm ik}^{\,(3)}(x)$ that are useful for high-precision calculations of 
benchmark processes in $ep$ and $pp$ scattering.

\vspace*{-1mm}
\section*{Acknowledgements}

\noindent
This work has been supported by the {\it European Research Council}$\,$ (ERC) 
Advanced Grant 320651, {\it HEPGAME} and the UK {\it Science \& Technology 
Facilities Council}$\,$ (STFC) grants ST/L000431/1 and ST/K502145/1. Part of 
our computations were carried out on the Dutch national e-infrastructure with 
the support of the SURF Cooperative and the PDP Group at Nikhef. We also are 
grateful for the opportunity to use a substantial part of the {\tt ulgqcd} 
computer cluster in Liverpool which was funded by the STFC grant ST/H008837/1.
A.V. would like to thank J. Bl\"umlein and A. Grozin for useful discussions 
during the workshop.

\end{document}